%% file: 00_Main.tex
\documentclass[sigconf]{acmart}

\usepackage{graphicx}
   \graphicspath{{./Figures/}}

\usepackage{amsmath}    
\usepackage{algorithm}
\usepackage{algorithmic}
\usepackage{multirow}
\usepackage{svg}
\usepackage{color}

\DeclareGraphicsExtensions{.pdf,.jpeg,.png,.fig, .emf}
   \usepackage{subfigure}
    \usepackage{epstopdf}
    \usepackage{epsfig}
    \usepackage{subfigure}
    \usepackage{epstopdf}
    \usepackage{epsfig}

\newcommand{\eat}[1]{}
\copyrightyear{2024}
\acmYear{2024}
\setcopyright{acmlicensed}\acmConference[DAC '24]{61st ACM/IEEE Design Automation Conference}{June 23--27, 2024}{San Francisco, CA, USA}
\acmBooktitle{61st ACM/IEEE Design Automation Conference (DAC '24), June 23--27, 2024, San Francisco, CA, USA}
\acmDOI{10.1145/3649329.3655990}
\acmISBN{979-8-4007-0601-1/24/06}

\begin{document}
%
\title{\vspace{-10mm} Energy Efficient Dual Designs of FeFET-Based Analog In-Memory Computing with Inherent Shift-Add Capability }

 \author{\small{}
         Zeyu Yang$^1$,
         Qingrong Huang$^1$,
         Yu Qian$^1$,
         Kai Ni$^2$,
         Thomas Kämpfe$^3$ and Xunzhao Yin$^{1,4*}$ 
          \\$^1$Zhejiang University, China
          $^2$University of Notre Dame, USA
          $^3$Fraunhofer IPMS, Germany
          \vspace{-1ex}
          \\$^4$Key Laboratory of CS\&AUS of Zhejiang Province, China 
          $^*$Corresponding author email: xzyin1@zju.edu.cn
}

\renewcommand{\bibfont}{\scriptsize}

\let\oldbibliography\thebibliography
\renewcommand{\thebibliography}[1]{\oldbibliography{#1}
\setlength{\itemsep}{-0.5pt}} 



\begin{abstract}
\vspace{-1ex}

In-memory computing (IMC) architecture emerges as a promising paradigm, improving the energy efficiency of multiply-and-accumulate (MAC) operations within deep neural networks (DNNs) by integrating the parallel computations within the memory arrays. 
Various high-precision analog IMC array designs have been developed based on both SRAM and emerging non-volatile memories (NVMs).
These designs perform MAC operations of partial input and weight, with the corresponding partial products then fed into shift-add circuitry to produce the final MAC results.
However, existing works often present intricate shift-add process for weight.
The traditional digital shift-add process is limited in throughput due to time-multiplexing of ADCs,  and advancing the shift-add process to the analog domain necessitates customized circuit implementations, resulting in  compromises in energy and area efficiency.
Furthermore, the joint optimization of the partial MAC operations and the weight shift-add process is rarely explored.
In this paper, we propose  novel, energy efficient dual designs of ferroelectric FET (FeFET) based high precision analog IMC featuring inherent shift-add capability.
We introduce a FeFET based IMC paradigm that performs partial MAC in each column, and inherently integrates the shift-add process for 4-bit weights by leveraging FeFET's analog storage characteristics.
\eat{This effectively eliminates the need for additional dedicated shift-add circuitry in multi-bit weight processing.}
This paradigm 
supports both 2's complement mode (2CM) and non-2's complement mode (N2CM) MAC, thereby offering flexible support for 4-/8-bit weight data in 2's complement format.
Building upon this paradigm, we propose novel FeFET based dual designs, CurFe for the current mode and ChgFe for the charge mode, to accommodate the high precision analog domain IMC architecture.
Evaluation results at circuit and system levels indicate that the circuit/system-level energy efficiency of the proposed FeFET-based analog IMC is 1.56$\times$/1.37$\times$ higher when compared to the state-of-the-art analog IMC designs.

\end{abstract}


\maketitle
\pagestyle{empty}

%

\input{01_Introduction}

\input{02_Background}
\input{03_Dual_Designs}

\input{04_Evaluation}

\input{05_Conclusion}




\vspace{-2.5ex}
\begin{acks}
\vspace{-1.5ex}

This work was supported in part by NSFC (92164203, 62104213) and SGC Cooperation Project (Grant No. M-0612).

\end{acks}

\vspace{-2.5ex}



%
\bibliographystyle{ieeetr}

\bibliography{bib}

\end{document}

%% file: 01_Introduction.tex
\vspace{-3ex}
\section{Introduction}
\label{sec:intro}
\vspace{-1ex}

Over the past decade, deep neural networks (DNNs) have 
become crucial in artificial intelligence (AI), particularly in domains such as image recognition, speech recognition, and dynamic monitoring \cite{lecun2015deep}. 
However, with technological advancements and  exponentially growing  data volumes, the computational and storage demands of DNNs have risen sharply. 
As a result, the significant data movement between memory and processing units has led to the "memory wall" bottleneck in the
conventional Von Neumann architecture.
In-memory computing (IMC) architecture, which integrates computational functions within memory, 
aims to address this  
by mitigating the 
extensive data movement. 
Its natural parallel processing capabilities are well-suited for 
efficient execution of multiply-and-accumulate (MAC) operations, which are essential to DNNs \cite{yan2022computing, verma2019memory, yan2022swim}.

In the analog IMC field, many SRAM-based designs have been proposed
\cite{biswas2018conv, si201924, dong202015, si202015, yue202014, su202116}. 
Nevertheless, the inherent binary storage nature of SRAM complicates multi-bit IMC circuitry design,
and the large SRAM cell size and standby power limit area and energy efficiency.  
In contrast, emerging non-volatile memories (NVMs) like resistive RAM (ReRAM), magnetic RAM (MRAM), 
and ferroelectric FET (FeFET), are gaining attention
due to their compact structure and near-zero standby power \cite{zhuo2022design, huang2023fefet, sebastian2020memory, xue202116, hung2021four, hung20228, soliman2020ultra, yin2022ferroelectric, saito2021analog, huang2021computing, yin2024ferroelectric, liu2023reconfigurable, xu2023challenges}. 
FeFET, in particular, is promising for building efficient IMC design to store weights and perform MAC operations in DNN inference due to multi-level cell (MLC) , high ON/OFF ratio, and  three-terminal read/write separation  \cite{li2020scalable, soliman2023first, hu2021memory, yin2020fecam, shou2023see, yin2023ultracompact}.

In high-precision IMC designs,  multi-bit MAC dataflow typically represents
each fixed-point $n$-bit weight value with $n$ cells in  adjacent columns. 
It thereby combines $n$ outputs of weighted partial sums based on different significant weight bits using peripheral circuits, a process called the shift-add process for weight.
Currently, there are two  shift-add process categories
\cite{jiang2021analog}. The first,  "digital shift-add", 
directs the partial MAC value (pMACV) of the selected column through a Multiplexer (MUX) to an analog-to-digital converter (ADC), which converts pMACV to digital form in a single cycle.
Over $n$ sequential cycles, these $n$ digital values are summed  on
bit significance via a digital shift-add circuit. 
This process is slow due to 
time-multiplexing, and consumes additional area overhead  for components like Multipliers  \cite{hur2022nonvolatile}. 
The alternative process,  "analog shift-add", 
relocates the shift-add process for weight before analog-to-digital conversion. This allows parallel generation of  pMACV results for each column, followed by their weighted summation in the analog domain. 
\eat{
Various analog shift-add circuit modules have been developed to accommodate diverse  analog domain based IMC implementations \cite{si201924, dong202015, yue202014}, but they invariably increase 
area overhead. 
Furthermore, the separation of 
the shift-add modules from the IMC array performing parallel MAC operations has left  potential for better integration of these implementations. \textcolor{red}{This sentence needs further clarification}
}
Several analog shift-add circuit modules have been developed to support various analog domain based IMC implementations \cite{si201924, dong202015, yue202014}. However, these solutions still involve significant energy  and area overheads.
Furthermore, both of the above shift-add circuit modules are separate from the array implementing the partial MAC operations, indicating a substantial opportunity for 
joint optimization of these two processes.

In this paper, we propose novel dual designs for  FeFET-based high precision analog IMC with inherent shift-add capability. 
Unlike previous approaches, we leverage FeFET's analog storage property to establish a novel FeFET-based IMC array paradigm that not only has partial MAC ability for each column, but also inherently integrates the shift-add process for 4-bit weights. This effectively eliminates the need for extra  dedicated shift-add circuits in multi-bit weight processing.
This FeFET based IMC paradigm supports both 2's/non-2's complement mode (2CM/N2CM) MAC
for 4-/8-bit weight data in 2's complement format. Building upon this paradigm, 
we propose  two designs for
prevalent analog domain IMC architectures: the current mode (CurFe) and the charge mode (ChgFe) architectures. 
The CurFe design uses 1nFeFET1R cells with binary-weighted conductance values, and a trans-impedance amplifier (TIA)  for both the MAC and shift-add processes. 
ChgFe  utilizes  single-level cell (SLC) 1pFeFET and  MLC 1nFeFET 
to store the sign bit and remaining bits, respectively, enabling inherent shift-add operations through charge sharing among capacitors.
SPICE and system-level simulation results suggest that our proposed FeFET-based analog IMC is 1.56$\times$/1.37$\times$ more energy efficient at the circuit/system-level compared to the state-of-the-art analog IMC designs.

The rest  is organized as follows: 
Section \ref{sec:background} reviews the basics and 
relevant prior works. Section \ref{sec:designs} introduces our proposed IMC dual designs, i.e., CurFe and ChgFe. Section \ref{sec:eval} presents  evaluation results. Section \ref{sec:conclusion} concludes the paper.

%% file: 02_Background.tex
\vspace{-3ex}
\section{Background}
\label{sec:background}
\vspace{-0.5ex}
Here we review the FeFET device, analog IMC  and related works.

\begin{figure}[!t]
  \centering
  \includegraphics[width=1\columnwidth]{Figures/devices.pdf}
  \vspace{-5ex}
  \caption{{\bf (a)}/{\bf (b)} Structure of nFeFET/pFeFET. {\bf (c)} Measured  $I_{D}$-$V_{G}$ characteristics with MLC $V_{th}$ states of a fabricated nFeFET.}
  \label{fig:devices}
  \vspace{-6ex}
\end{figure}

\vspace{-2.5ex}
\subsection{FeFET Basics}
\label{subsec:FeFET_Basics}
\vspace{-1ex}

FeFET,  a three-terminal NVM device, 
is widely used in IMC designs due to its 
CMOS compatibility,  high ON/OFF ratio, and compact structure. 
Most research focuses on nFeFET, similar to a standard nMOS transistor but featuring a thick doped HfO2 ferroelectric (FE) layer on its gate, as shown in Fig. \ref{fig:devices}(a).  
Measurements on fabricated nFeFET devices as shown in Fig. \ref{fig:devices}(c) have indicated 
that nFeFETs can be programmed to exhibit MLC threshold  voltage $V_{TH}$ states  
by applying different write pulses at gate. \eat{\cite{yin2020fecam}.} 
This capability
enables non-volatile storage of information  within an 1nFeFET cell. 
In parallel, the pFeFET device, based on pMOS transistor technology (Fig. \ref{fig:devices}(b)), has been experimentally demonstrated to exhibit similar switching behavior to nFeFET \cite{kleimaier2021demonstration}. 
Furthermore, the 1pFeFET cell 
has been validated 
as a programmable synapse \cite{thomann2022all}, demonstrating its usefulness in IMC applications.

\vspace{-3ex}
\subsection{Analog IMC preliminaries} 
\label{subsec:existIMC}
\vspace{-0.5ex}

Numerous SRAM-based analog IMC designs have been proposed \cite{biswas2018conv, si201924, dong202015, si202015, yue202014, su202116}.
On the one hand, 
these designs explore a variety of 
cell structures, including 8T-SRAM \cite{si201924, dong202015, yue202014}, 10T-SRAM \cite{biswas2018conv}, and Block-wise 6T SRAM \cite{si202015, su202116}, among others. 
In terms of implementation methods in analog domain, there are primarily two modes: 
current mode and charge mode. 
Current mode aggregates  currents from multiple computing units  to obtain pMACV \cite{si201924, si202015, yue202014}, while charge mode represents multiplication results as charges \cite{biswas2018conv, dong202015, su202116}.
Additionally, analog IMC architectures based on various NVMs have emerged.
ReRAM-based designs are studied for 8-bit precision inference implementation \cite{xue202116, hung2021four, hung20228}.
FeFET has recently gained attention as a potential candidate for analog IMC, thanks to its three-terminal structure, high ON/OFF ratio, and compact structure. However, most existing FeFET-based IMC designs are limited to storing only binary states \cite{soliman2020ultra, saito2021analog}, 
not  fully utilizing the MLC states of FeFET. 
Recently, Soliman et al. \cite{soliman2023first} proposed an IMC design using a MLC FeFET for 2-bit multiplication operations in an 1nFeFET1R cell.
Nevertheless, current limitations in the precision that can be stored in a single FeFET make extending this design to high-precision IMC, such as 8-bit, a notable  challenge.

\vspace{-3ex}
\subsection{Related Works and Motivation} 
\label{subsec:existIMC}
\vspace{-1ex}

Storing an $n$-bit weight in a single IMC cell is typically challenging, requiring the use of multiple cells in adjacent columns. 
As a result, this necessitates combining the pMACVs from various columns 
based on their   weight bit significance through subsequent peripheral circuitry, a process known as the shift-add process for weight. 
The conventional "digital shift-add" process is  executed after ADC conversion with a digital circuit module.
However, due to the considerable ADC overhead, multiple columns have to share an ADC through time-multiplexing, facilitated by a MUX, thus 
demanding multiple clock cycles to complete a MAC operation. 
To optimize this process, several  works adopt the "analog shift-add" approach \cite{si201924, dong202015, yue202014}, 
where the shift-add operation occurs  before the ADC conversion in the analog domain. 
This method allows the ADC to directly generate the MAC result 
with embedded weight significance, leading to a substantial throughput improvement.
Si et al. \cite{si201924} proposed a 2's complement weight mapping scheme with a processing unit to achieve "analog shift-add" for 5-bit weights, 
it requires additional proportional capacitors, resulting in extra area overhead. 
Dong et al. \cite{dong202015} employed binary-weighted computation capacitors in 4-bit Flash ADCs for charge sharing and  column-wise pMACVs combining for 4-bit precision, 
but the scalability  to higher weight bit precision is challenging due to the significant capacitance difference between the least significant bit (LSB) capacitor and the most significant bit (MSB) capacitor. 
Yue et al. \cite{yue202014} addressed the above issues and designed an "analog shift-add" based ADC that flexibly supports both 2CM/N2CM for signed/unsigned 4-bit weight,
thus allowing for 8-bit signed weight MAC operations 
using 2CM for the high 4-bit and N2CM for the low 4-bit. 
For example, an $m$-bit unsigned input $X$
(where $m$=2 in \cite{yue202014}) 
and an 8-bit signed weight $Y$ in 2's complement format. 
The multiplication is segmented into two parts corresponding to the 2CM/N2CM ADCs:
\vspace{-1.3ex}
\begin{equation}
\label{equ:xy}
    X=\sum_{i=0}^{m-1}{x_i}2^{i}, Y=(-{y_7}2^{7}+\sum_{j=4}^{6}{y_j}2^{j})+(\sum_{j=0}^{3}{y_j}2^{j})
\end{equation}
\vspace{-2ex}
\begin{equation}
\label{equ:mul}
    XY=(-{y_7}\sum_{i=0}^{m-1}{x_i}2^{i+7}+\sum_{i=0}^{m-1}\sum_{j=4}^{6}{x_i}{y_j}2^{i+j})+(\sum_{i=0}^{m-1}\sum_{j=0}^{3}{x_i}{y_j}2^{i+j})
\end{equation}
\vspace{-0.4ex}
When combining  the accumulation module, such MAC operation support can be extended to 2-/4-/6-/8-bit inputs and 4-/8-bit weights. 
However, like the approach in \cite{si201924}, extra binary-weighted capacitors are also required.
Besides, the separation of both types of the shift-add process for weight from the multi-bit MAC operation suggests 
potential for further integration of these processes. 
In this paper, we propose energy efficient dual designs for  FeFET-based analog IMC with inherent shift-add capability, 
aiming to integrate the shift-add process for weight with the partial MAC operation in the IMC array through FeFET's analog storage characteristics.

%% file: 03_Dual_Designs.tex
\vspace{-2.2ex}
\section{Proposed Analog IMC Dual Designs}
\label{sec:designs}
\vspace{-1ex}
Here we introduce the dual designs of analog IMC paradigm that utilizes the adjustable analog characteristics of FeFET  cells 
for storing multi-bit weights.
Our current mode design uses 1nFeFET1R cells with varying resistance to conduct weighted currents corresponding to weighted bits.
Meanwhile, our charge mode design conducts weighted currents by programming different $V_{TH}$ to 1nFeFET cells.
\vspace{-5ex}
\subsection{Current Mode FeFET-Based IMC}
\label{sec:currentmode}
\vspace{-1ex}

\begin{figure}[!t]
  \centering
  \includegraphics[width=1\columnwidth]{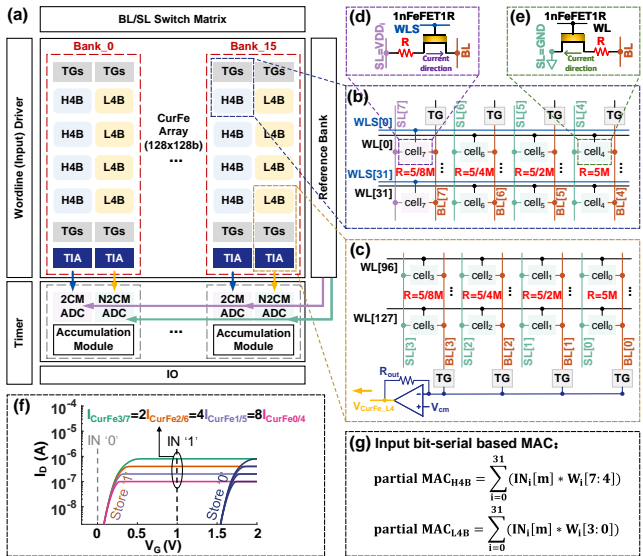}
  \vspace{-5.8ex}
  \caption{{\bf (a)} Structure of the proposed CurFe architecture. {\bf (b)} Structure of H4B with TGs. {\bf (c)} Structure of L4B with TGs and TIA. {\bf (d)} 1nFeFFET1R structure for $cell_{7}$. {\bf (e)} 1nFeFFET1R structure for $cell_{0}$-$cell_{6}$. {\bf (f)} Id-Vg curves of $cell_{0}$-$cell_{7}$. {\bf (g)} Input bit-serial based MAC with two parts in H4B and L4B.}
  \label{fig:arch_CurFe}
  \vspace{-5.2ex}
\end{figure}

This section introduces the design of the current mode FeFET-based IMC, referred to as CurFe. Fig. \ref{fig:arch_CurFe}(a) illustrates the overall architecture, comprising a wordline (input) driver, a BL/SL switch matrix, a  128x128b array based on 1nFeFET1R cells, a reference bank, 16 2CM ADCs, 16 N2CM ADCs, 16 accumulation modules, and other peripheral circuits.
The 
array is divided into 16 banks, each containing 4 high 4-bit blocks (H4Bs), 4 low 4-bit blocks (L4Bs), 16 transmission gates (TGs), and 2 TIAs. 
As depicted in Fig. \ref{fig:arch_CurFe}(b), each H4B consists of 32 rows and 4 columns of 1nFeFER1R cells, storing 32 4-bit signed weight data. 
Similarly, each L4B, as shown in Fig. \ref{fig:arch_CurFe}(c),  has a 32-row, 4-column configuration of 1nFeFER1R cells for storing 32 4-bit unsigned weight data.
Note that cells labeled as $cell_{7}$ (Fig. \ref{fig:arch_CurFe}(d)) in the same row of H4Bs share a common wordline, WLS, while other cells depicted in Fig. \ref{fig:arch_CurFe}(e) in the same row of H4Bs/L4Bs share another wordline, WL.
All cells in the same column share identical sourcelines (SL) and bitlines (BL). 
The 1nFeFET1R cell 
incorporates a drain resistance to ensure
that the ON state current is limited by the resistance, significantly reducing 
ON state current variation \cite{soliman2020ultra}. 
The drain resistances in $cell_{0}$, $cell_{1}$, $cell_{2}$, and $cell_{3}$  are set to 5M, 5/2M, 5/4M, and 5/8M, respectively, and the same  configuration is applied from $cell_{4}$ to $cell_{7}$. 
Each SLC 1nFeFET1R cell can be written to either a low $V_{TH}$ state or a high $V_{TH}$ state, corresponding to 1-bit weighted "1" and "0" values, respectively.

\begin{figure}[!t]
  \centering
  \includegraphics[width=1\columnwidth]{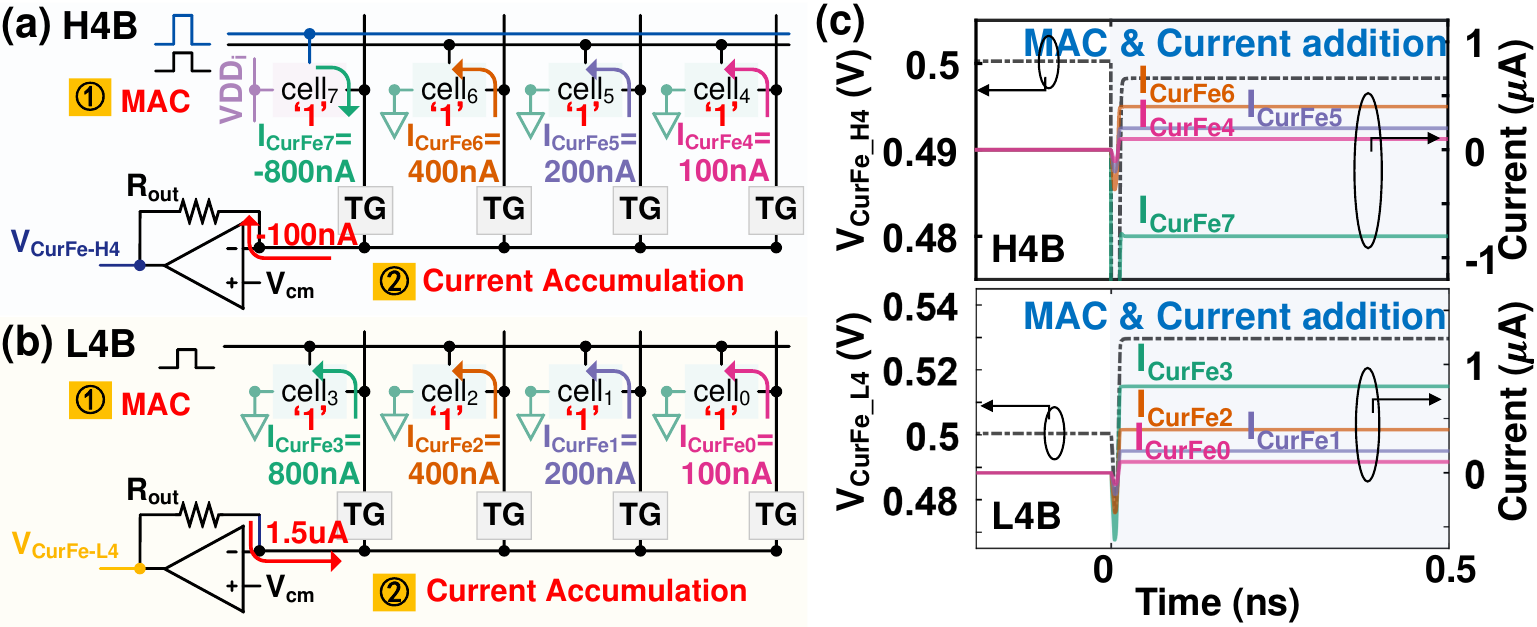}
  \vspace{-5.8ex}
  \caption{Multiplication example of an 1-bit input and 8-bit signed weight in CurFe. The 8-bit weight is divided into  {\bf (a)} high 4-bit and {\bf (b)} low 4-bit parts in H4B and L4B, respectively. {\bf (c)} Transient simulation waveforms of  this operation.}
  \label{fig:illus_CurFe}
  \vspace{-5ex}
\end{figure}

The proposed CurFe architecture 
supports 1-8 bit unsigned inputs and 4-/8-bit signed weights in 2's complement format, enabled by ADCs, accumulation modules, and external control. 
Multi-bit input data processing occurs in bit-serial mode, with each 8-bit weight divided into high 4-bit and low 4-bit segments stored in adjacent columns of H4B and L4B, respectively,
as expressed in 
Fig. \ref{fig:arch_CurFe}(g).
To perform 32 accumulations for the multiplication of 1-bit input and 8-bit weight, a set of H4B and L4B is activated in each bank 
for each serial input bit. 
Each 1-bit input data is  applied to  corresponding WL and WLS through the wordline driver. 
Controlled by multiple TGs, BL[4]-BL[7]/BL[0]-BL[3] are connected to  the TIA's  inverting input  in H4B/L4B. 
The voltage at this node approximates the bias voltage $V_{cm}$ (0.5V) at the noninverting input.
With SL[7]  set to $VDD_{i}$ (1V), and other SLs  grounded, 
the 1nFeFET1R cell can perform the multiplication of 1-bit input and 1-bit weight. 
As shown in Fig. \ref{fig:arch_CurFe}(f), thanks to the varied drain resistances, the ON state currents associated with $cell_{0}$-$cell_{3}$ (denoted as $I_{CurFe0}$-$I_{CurFe3}$) and $cell_{4}$-$cell_{7}$ (denoted as $I_{CurFe4}$-$I_{CurFe7}$) follow a binary-weighted pattern, with
the direction of $I_{CurFe7}$ being opposite to the others. 
Consequently, the two TIAs within each bank  collectively accumulate all ON state currents from the activated H4B/L4B to produce the output voltages $V_{CurFe-H4}$/$V_{CurFe-L4}$:
\vspace{-1ex}
\begin{equation}
\label{equ:CurFe_H4}
\begin{aligned}
    V_{CurFe-H4}=V_{cm}+(\sum I_{CurFe7}+\sum I_{CurFe6}\\+\sum I_{CurFe5}+\sum I_{CurFe4})*R_{out}
\end{aligned}
\end{equation}
\vspace{-2ex}
\begin{equation}
\label{equ:CurFe_L4}
\begin{aligned}
    V_{CurFe-L4}=V_{cm}+
    (\sum I_{CurFe3}+\sum I_{CurFe2}\\+\sum I_{CurFe1}+\sum I_{CurFe0})*R_{out}
\end{aligned}
\end{equation}
where $R_{out}$ is the feedback resistor on the TIA. 
In essence, 
Eq. \eqref{equ:CurFe_H4} and \eqref{equ:CurFe_L4} imply the integration of the 1-bit partial MAC operation and the shift-add process for 4-bit signed/unsigned weight in 2CM/N2CM, repsectively.
\eat{Fig. \ref{fig:illus_CurFe}(a) and (b) illustrate an  example of  multiplying  an 1-bit input '1' with an 8-bit weight "11111111" in CurFe,
with the resultant accumulated currents on the TIAs in H4B and L4B being -100nA and 1.5$\mu$A, respectively, as shown in 
the transient waveform  in Fig. \ref{fig:illus_CurFe}(c).}
Fig. \ref{fig:illus_CurFe}(a) and (b) illustrate an example of multiplying an 1-bit input '1' with an 8-bit weight "11111111" in CurFe, while none of the other rows in this H4B/L4B  are enabled. The resultant accumulated currents on the TIAs in H4B and L4B are -100nA and 1.5$\mu$A, respectively. Thus, the output voltages are obtained as shown in the transient waveform presented in Fig. \ref{fig:illus_CurFe}(c).

The SAR-ADC, as proposed by Yue et al. \cite{yue202014}, flexibly supports both 2CM/N2CM for signed/unsigned weights. Our design  employs this  ADC type to operate in 2CM/N2CM, converting the analog pMACVs corresponding to the high 4-bit and low 4-bit parts of the signed 8-bit weights into their respective digital forms. 
The reference voltages  for 2CM and N2CM ADCs are  internally generated by the reference bank,
an approach  previously employed in \cite{si201924, si202015, su202116}.
After the ADC conversion, the MAC operation for 8-bit weights is achieved by combining the results of the 2CM ADC and N2CM ADC within the same bank, processed in the accumulation module. 
To support input precision exceeding 1-bit,
the input bit-serial based MAC process described above is iterated, and  the shift-add operation for input is completed in the accumulation module.

\vspace{-2.5ex}
\subsection{Charge Mode FeFET-Based IMC}
\label{sec:chargemode}
\vspace{-0.5ex}

\begin{figure}[!t]
  \centering
  \includegraphics[width=1\columnwidth]{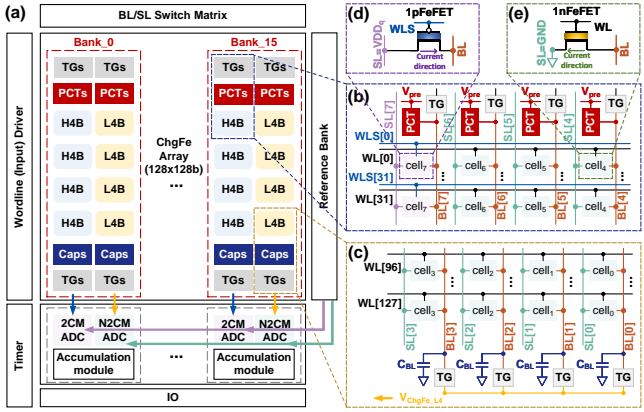}
  \vspace{-5.8ex}
  \caption{{\bf (a)} Structure of the proposed ChgFe architecture. {\bf (b)} Structure of H4B with PCTs ans TGs. {\bf (c)} Structure of L4B with capacitors and TGs. {\bf (d)} 1pFeFET structure for $cell_{7}$. {\bf (e)} 1nFeFET structure for $cell_{0}$-$cell_{6}$.}
  \label{fig:arch_ChgFe}
  \vspace{-3.8ex}
\end{figure}

\begin{figure}[!t]
  \centering
  \includegraphics[width=1\columnwidth]{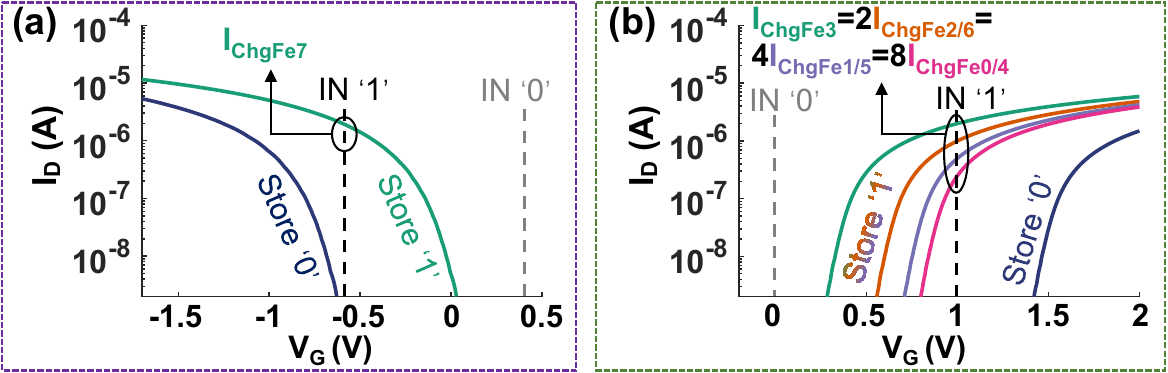}
  \vspace{-5.8ex}
  \caption{{\bf (a)} Id-Vg curves of $cell_{7}$ in ChgFe. {\bf (b)} Id-Vg curves of $cell_{0}$-$cell_{6}$ in ChgFe.}
  \label{fig:1FeFET_Id-Vg}
  \vspace{-5.5ex}
\end{figure}

This section describes the design of charge mode FeFET-based IMC, denoted as ChgFe. As depicted in Fig. \ref{fig:arch_ChgFe}(a), the  architecture of ChgFe is similar to  CurFe and includes a wordline driver, a BL/SL switch matrix, a 128x128b array, a reference bank, 16 2CM ADCs, 16 N2CM ADCs, 16 accumulation modules, and other peripheral circuits. 
However, in the 128x128b ChgFe array, each $cell_{7}$ features a 1pFeFET cell (Fig. \ref{fig:arch_ChgFe}(d)), while the other cells use 1nFeFET cells (Fig. \ref{fig:arch_ChgFe}(e)). 
Additionally, as illustrated in Fig. \ref{fig:arch_ChgFe}(b), each BL is associated with a pre-charged transistor (PCT) and a capacitor, replacing the TIA used in CurFe.
Notably, the $V_{TH}$ state positioned on the right side in Fig. \ref{fig:1FeFET_Id-Vg}(a) is designated as the high $V_{TH}$ state of pFeFET, 
indicating that the sign bit for the weight has a value of '1'.
To signify weight significance, as shown in Fig. \ref{fig:1FeFET_Id-Vg}(b),  the low $V_{TH}$ states (representing '1') of the 1nFeFET cells for distinct significant weight bits vary, and the ON current magnitude of the high $V_{TH}$ state of the 1pFeFET in $cell_{7}$ matches that of $cell_{3}$, creating a binary-weighted pattern for the ON state current of $cell_{0}$-$cell_{3}$ (denoted as $I_{ChgFe0}$-$I_{ChgFe3}$)/$cell_{4}$-$cell_{7}$ (denoted as $I_{ChgFe4}$-$I_{ChgFe7}$).

\begin{figure}[!t]
  \centering
  \includegraphics[width=1\columnwidth]{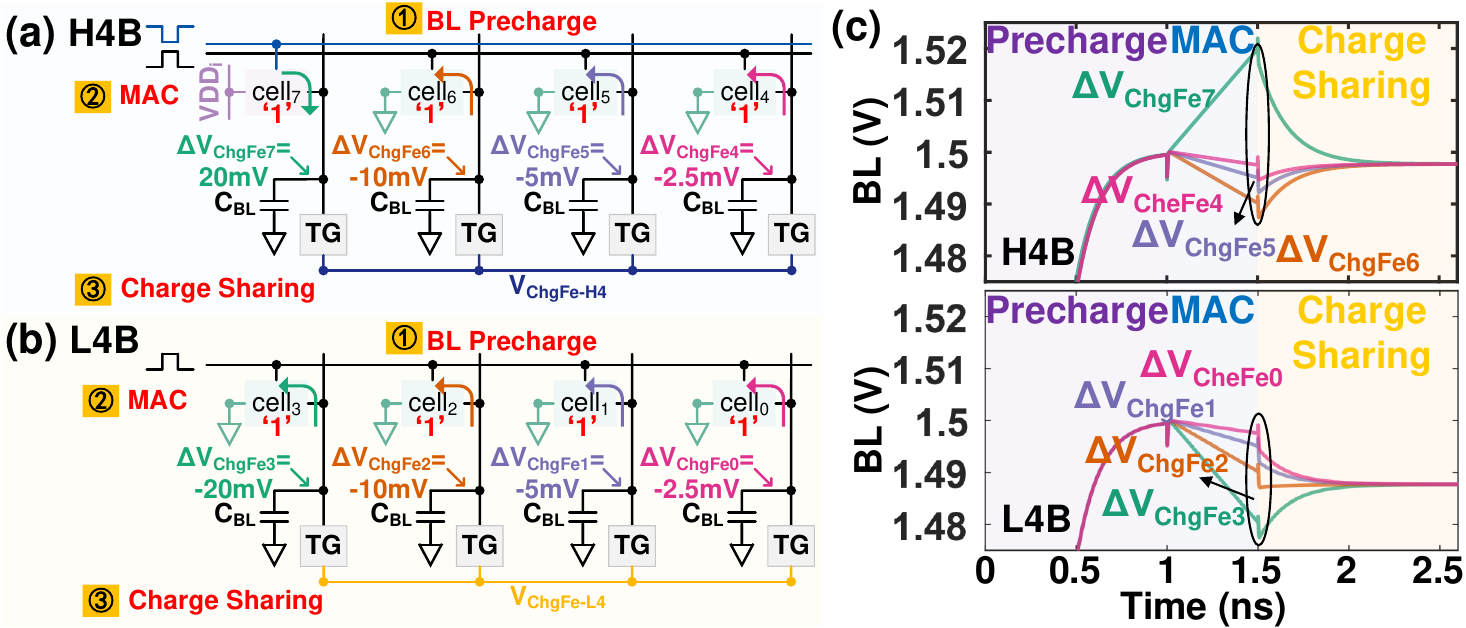}
  \vspace{-5.5ex}
  \caption{Multiplication example of an 1-bit input and 8-bit signed weight in ChgFe. The 8-bit weight is divided into  {\bf (a)} high 4-bit and {\bf (b)} low 4-bit parts in H4B and L4B, respectively. {\bf (c)} Transient simulation waveforms of this  operation.}
  \label{fig:illus_ChgFe}
  \vspace{-3ex}
\end{figure}

The proposed ChgFe performs 32 accumulations for an 1-bit input and an 8-bit weight in the charge domain. 
Within each bank, SL[7] is set to $VDD_{q}$, while  other SLs are grounded, and all TGs are  OFF. 
Initially, each BL capacitor is pre-charged to $V_{pre}$ (1.5V) within 1ns via PCT. 
32 1-bit input data are then applied to the corresponding WL and WLS via wordline drivers within  0.5ns.
In this stage, the MAC operation for 1-bit input and 1-bit weight is executed on each BL. Specifically, the capacitor on BL[7] is charged through activated $cell_{7}$s, while the capacitors on BL[0]-BL[6] are discharged through the respective activated $cell_{0}$s-$cell_{6}$s. Since all FeFETs operate in the saturation region, the ON state currents vary less with time, causing
the voltage changes in BL voltages (denoted as $\Delta V_{ChgFe0}$-$\Delta V_{ChgFe3}$ for $cell_{0}$-$cell_{3}$ and $\Delta V_{ChgFe4}$-$\Delta V_{ChgFe7}$ for $cell_{4}$-$cell_{7}$, respectively) 
following a binary-weighted pattern, with 
$\Delta V_{ChgFe7}$ being positive and  others  negative. 
Subsequently, 
controlled by multiple TGs, 
charge sharing operations between BL capacitors in H4B/L4B are performed to produce the output voltages $V_{ChgFe-H4}$/$V_{ChgFe-L4}$:
\vspace{-1ex}
\begin{equation}
\label{equ:ChgFe_H4}
\begin{aligned}
    V_{ChgFe-H4}=V_{pre}+(\sum\Delta V_{ChgFe7}+\sum\Delta V_{ChgFe6}\\+\sum\Delta V_{ChgFe5}+\sum\Delta V_{ChgFe4})/4
\end{aligned}
\end{equation}
\vspace{-1ex}
\begin{equation}
\label{equ:ChgFe_L4}
\begin{aligned}
    V_{ChgFe-L4}=V_{pre}+(\sum\Delta V_{ChgFe3}+\sum\Delta V_{ChgFe2}\\+\sum\Delta V_{ChgFe1}+\sum\Delta V_{ChgFe0})/4
\end{aligned}
\end{equation}
Hence, both the 1-bit partial MAC operation and the shift-add process for 4-bit weight in 2CM/N2CM depend on the same BL capacitors, eliminating the need for extra binary-weighted computation capacitors as seen in \cite{dong202015}.
Fig. \ref{fig:illus_ChgFe}(a) and (b) illustrate a 
multiplication scenario of an 1-bit input '1' and an 8-bit weight "11111111" in ChgFe, with no other rows in this H4B/L4B activated.
As shown in Fig. \ref{fig:illus_ChgFe}(c), each BL voltage experiences a slight drop at the beginning of the charge-sharing operation, but this does not affect linearity.

Similar to  CurFe,  ChgFe  can accommodate 1-8 bit unsigned inputs and 4-/8-bit signed weights in 2's complement format, aided by ADCs, accumulation modules, and external control. The details of this operation are similar to  those in CurFe and    are not repeated.

%% file: 04_Evaluation.tex
\vspace{-2.5ex}
\section{Validation and Evaluation}
\label{sec:eval}
\vspace{-1ex}

In this section, we validate and evaluate our proposed dual designs of FeFET-based analog IMC at both circuit level and system level, which are then compared 
with  state-of-the-art analog IMC designs.

\begin{figure}[!t]
  \centering
  \includegraphics[width=1\columnwidth]{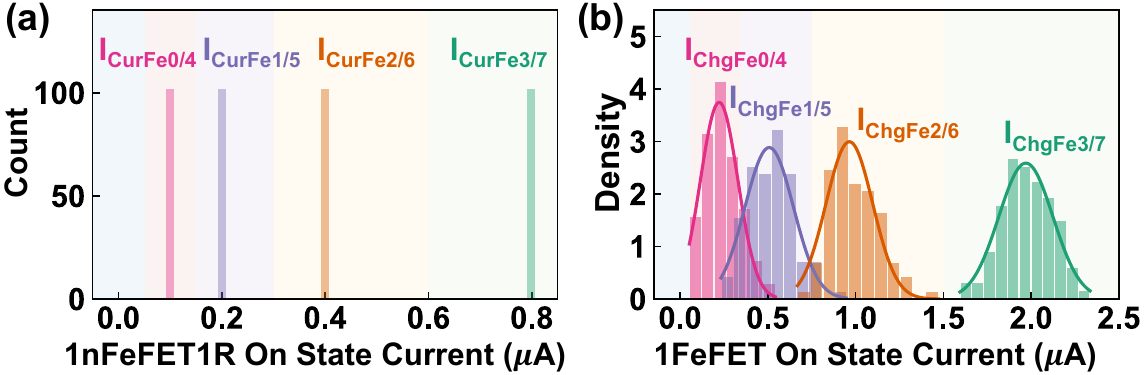}
  \vspace{-5ex}
  \caption{{\bf (a)} Current histogram of $I_{CurFe_{0}}$-$I_{CurFe_{7}}$ in CurFe. {\bf (b)} Current histogram of $I_{ChgFe_{0}}$-$I_{ChgFe_{7}}$ in ChgFe.} 
  \label{fig:DenandHis}
  \vspace{-3.5ex}
\end{figure}

\vspace{-2.5ex}
\subsection{Circuit-Level Validation and Evaluation}
\label{sec:results}
\vspace{-1ex}


\begin{figure}[!t]
  \centering
  \includegraphics[width=1\columnwidth]{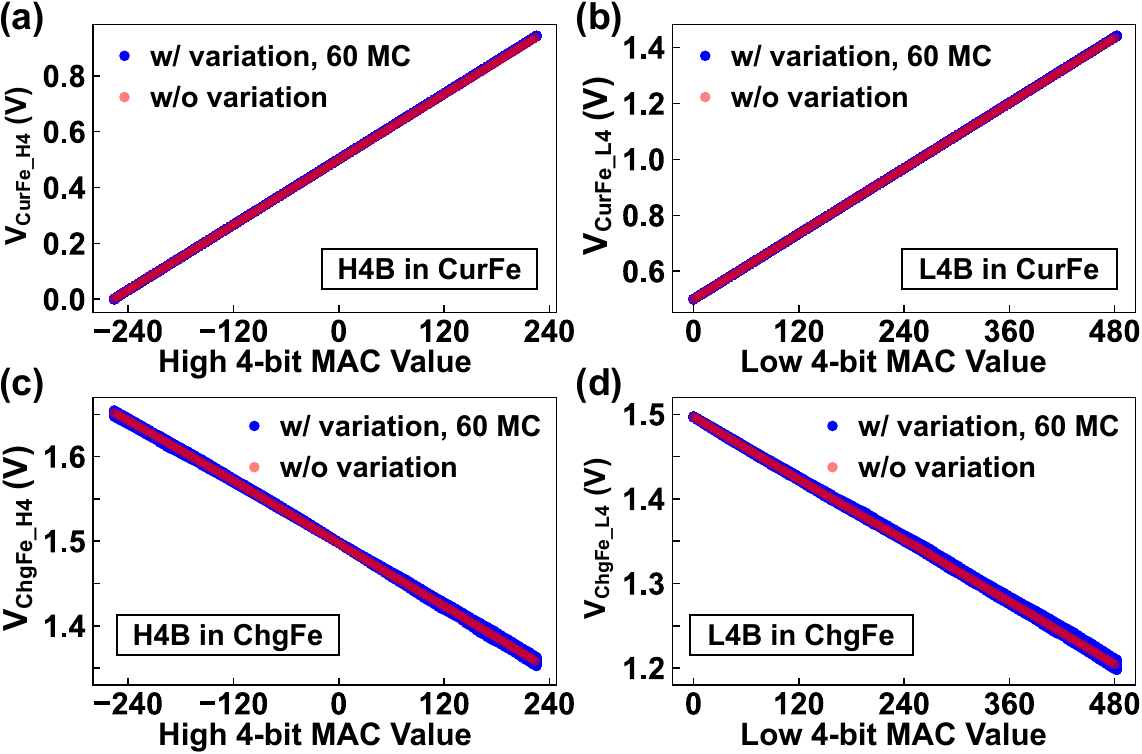}
  \vspace{-5ex}
  \caption{The MAC outputs for 32 accumulations
  of 1-bit input and 4-bit weight in {\bf (a)}/{\bf (c)} H4B and {\bf (b)}/{\bf (d)} L4B under CurFe/ChgFe.}
  \label{fig:MACout}
  \vspace{-3ex}
\end{figure}

\eat{
\begin{figure}[!t]
  \centering
  \includegraphics[width=1\columnwidth]{Figures/1nFeFET1R_Id-Vg.pdf}
  \vspace{-3ex}
  \caption{{\bf (a)} Id-Vg curves of 
  $cell_{0}$-$cell_{7}$ in CurFe, and {\bf (b)} the corresponding current histogram of $I_{CurFe_{0}}$-$I_{CurFe_{7}}$.} 
  \label{fig:1nFeFET1R_Id-Vg}
\end{figure}
}

\eat{
\begin{figure}[!t]
  \centering
  \includegraphics[width=1\columnwidth]{Figures/1nFeFET_Id-Vg.pdf}
  \vspace{-3ex}
  \caption{{\bf (a)} Id-Vg curves of 
  $cell_{0}$-$cell_{6}$ in ChgFe, and {\bf (b)} the corresponding current histogram of $I_{ChgFe_{0}}$-$I_{ChgFe_{6}}$.} 
  \label{fig:1nFeFET_Id-Vg}
\end{figure}
}

\eat{
\begin{figure}[!t]
  \centering
  \includegraphics[width=1\columnwidth]{Figures/1pFeFET_Id-Vg.pdf}
  \vspace{-3ex}
  \caption{{\bf (a)} Id-Vg curves of 
  $cell_{7}$ in ChgFe, and {\bf (b)} the corresponding current histogram of $I_{ChgFe_{7}}$.} 
  \label{fig:1pFeFET_Id-Vg}
\end{figure}
}

\eat{
\begin{figure}[!t]
  \centering
  \includegraphics[width=1\columnwidth]{Figures/circuitperfm.pdf}
  \vspace{-3ex}
  \caption{{\bf (a)} Average energy per MAC and {\bf (b)} average circuit-level energy efficiency for 32 accumulations with different input and weight precision in CurFe/ChrFe.}
  \label{fig:circuitperfm}
  \vspace{-4ex}
\end{figure}
}

\begin{figure}[!t]
  \centering
  \includegraphics[width=0.9\columnwidth]{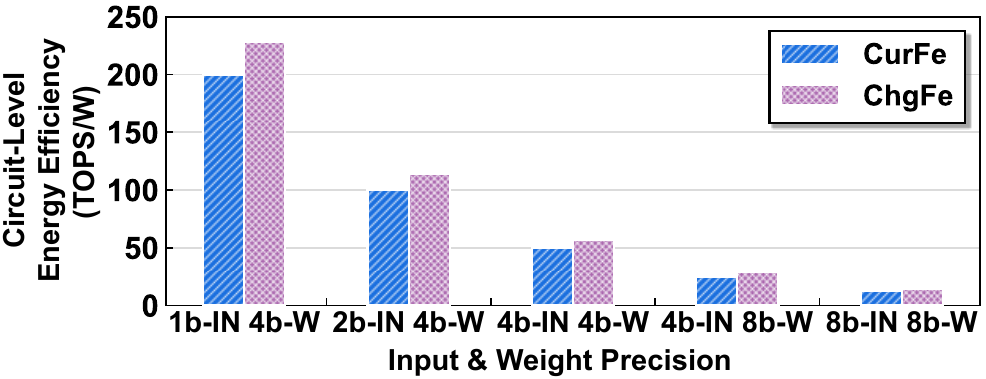}
  \vspace{-3ex}
  \caption{Average  energy efficiency for 32 accumulations with different input and weight precision in CurFe/ChrFe.}
  \label{fig:circuitperfm_ave}
  \vspace{-5ex}
\end{figure}

We conducted SPICE simulations on our proposed dual designs of FeFET-based analog IMC using the Cadence Spectre Simulator.
The simulations are based on the experimentally calibrated Preisach FeFET model \cite{ni2018circuit} and a commercial CMOS 40nm processing development kit.
The write method described in \cite{reis2019design} was adopted.
The BL capacitor in ChgFe was set to 50fF. 
FeFET devices are assumed to have  $V_{TH}$ variability with $\sigma$=40mV for each state, 
as per \cite{soliman2023first}.

We first  performed a Monte Carlo simulation to assess the impact of FeFET variation on different ON state currents. 
In the CurFe architecture, as shown in Fig. \ref{fig:DenandHis}(a), the drain resistance of the 1nFeFET1R cell significantly mitigates fluctuations in the ON state current. 
For ChgFe, to maintain adequate linearity in the charge domain, we employed the 1nFeFET and 1pFeFET cell structure.
The simulation results are presented in Fig. \ref{fig:DenandHis}(b).
We then exhaustively examined all possible input and weight combinations to generate complete MAC outputs for 32 accumulations of 1-bit input and 4-bit weight in H4B and L4B under both CurFe and ChgFe designs.
As illustrated in Fig. \ref{fig:MACout}, the results  exhibit good linearity for both CurFe and ChgFe.
By conducting 60 Monte Carlo simulations for each case, we observe the impact of FeFET variation on the output voltages, consistent with the effect on the currents in Figure 
\ref{fig:DenandHis}.
Furthermore, we accounted for different output  fluctuations in CurFe and ChgFe to assess the DNN inference accuracy in   Section \ref{sec:Benchmark}.

\eat{
In terms of energy consumption, as depicted in Fig. \ref{fig:circuitperfm}(a), we evaluated the average energy  per MAC for 32 accumulations involving different input and weight precision in both CurFe and ChgFe. 
As expected,  energy consumption per MAC  increases with precision.
Notably, the energy consumption in ChgFe is lower than  in CurFe at the same precision level.
This difference is attributed to the higher energy requirement for the TIA in CurFe compared to the energy nedded for precharging in ChgFe. 
Consequently, the average circuit-level energy efficiency exhibits an opposing trend, as illustrated in Fig. \ref{fig:circuitperfm} (b).
}
In terms of energy efficiency, as depicted in Fig. \ref{fig:circuitperfm_ave}, we evaluated the average circuit-level energy efficiency for 32 accumulations involving different input and weight precision in both CurFe and ChgFe. 
It is essential to highlight that $x$b-IN/$y$b-W represents the $x$-bit input/$y$-bit weight precision in Fig. \ref{fig:circuitperfm_ave} and subsequent figures.
This evaluation was conducted with a 5-bit ADC precision setting. A detailed analysis of ADC precision will be presented in Section \ref{sec:Benchmark}.
As expected,  energy efficiency decreases with input/weight precision.
Notably, the energy efficiency in CurFe is lower than that in ChgFe at the same precision level.
This difference is attributed to the higher energy requirement for the TIA in CurFe compared to the energy needed for precharging in ChgFe.

\begin{figure}[!t]
  \centering
  \includegraphics[width=1\columnwidth]{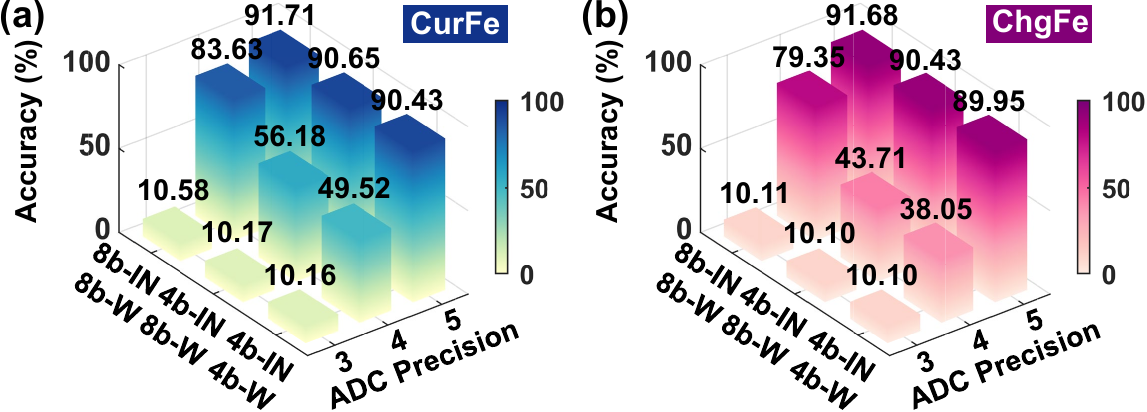}
  \vspace{-5ex}
  \caption{Impact of ADC resolution and input/weight precision on accuracy for CIFAR10 dataset in {\bf (a)} CurFe and {\bf (b)} ChgFe architectures.}
  \label{fig:systemacc}
  \vspace{-3.5ex}
\end{figure}

\begin{figure}[!t]
  \centering
  \includegraphics[width=1\columnwidth]{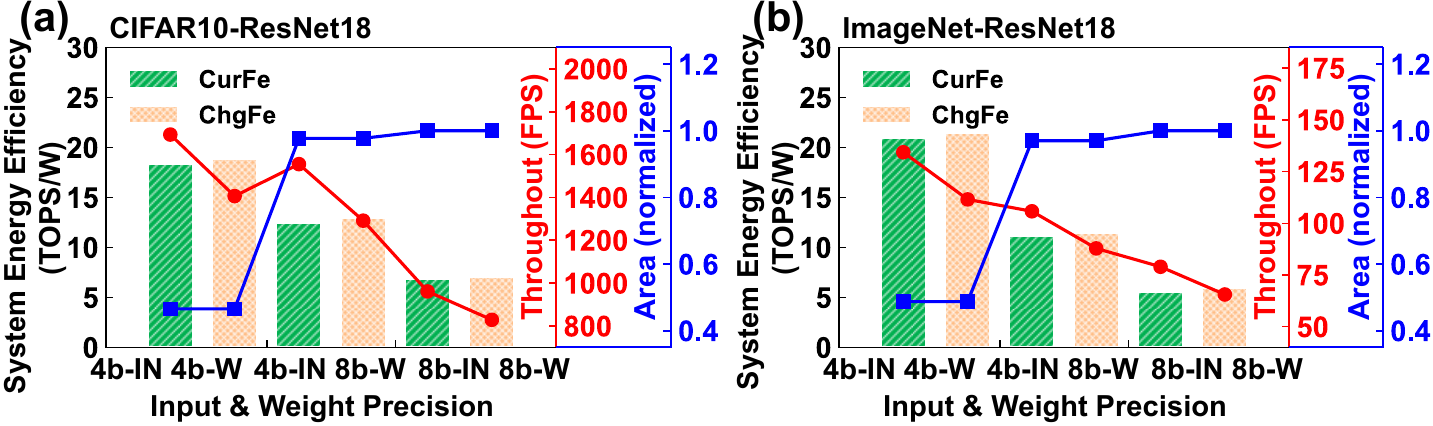}
  \vspace{-6ex}
  \caption{System performance of the proposed CurFe/ChgFe architectures with different input and weight precision on ResNet18 for {\bf (a)} CIFAR10 and {\bf (b)} ImageNet datasets.}
  \label{fig:systemperfm}
  \vspace{-5.5ex}
\end{figure}

\vspace{-2.5ex}
\subsection{Benchmark Results and Discussion}
\label{sec:Benchmark}
\vspace{-1ex}

\begin{figure*}[!t]
  \centering
  \includegraphics[width=1.60\columnwidth]{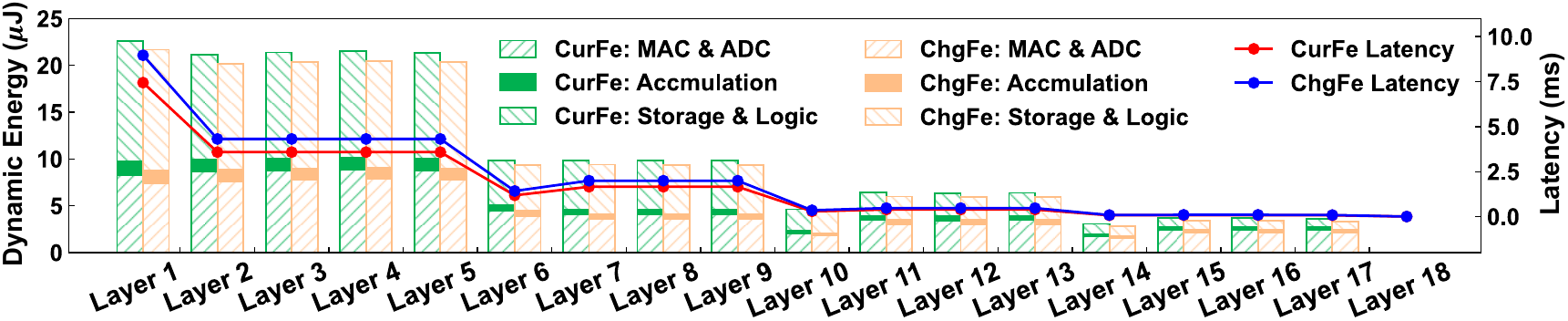}
  \vspace{-3ex}
  \caption{Breakdown of dynamic energy consumption and latency for each layer of ResNet18 using ImageNet dataset with 4-bit input and 4-bit weight precision in CurFe/ChgFe architectures.}
  \label{fig:systemeachlayer}
  \vspace{-3.5ex}
\end{figure*}

\begin{table*}
\caption{Comparison with the state-of-the-art analog IMC designs}
\label{table:evaluation}
\vspace{-3ex}
\centering
\resizebox{\linewidth}{!}{
\begin{tabular}{|c|c|c|c|c|c|c|c|c|}
\hline\hline
Reference                                                                                           & \cite{si202015}                                                           & \cite{yue202014}                                                                              & \cite{su202116}                                                   & \cite{xue202116}                                                  & \cite{hung2021four}                                               & \cite{hung20228}                                                  & CurFe                                                                   & ChgFe                                                              \\ \hline
Technology                                                                                          & CMOS                                                                      & CMOS                                                                                          & CMOS                                                              & ReRAM                                                             & ReRAM                                                             & ReRAM                                                             & FeFET                                                                   & FeFET                                                              \\ \hline
Cell Type                                                                                           & 6T-SRAM+LLC                                                               & 8T-SRAM                                                                                       & 6T-SRAM+LMC                                                       & 1T1R                                                              & 1T1R                                                              & 1T1R                                                              & 1nFeFET1R                                                               & 1nFeFET/1pFeFET                                                    \\ \hline
Node                                                                                                & 28nm                                                                      & 65nm                                                                                          & 28nm                                                              & 22nm                                                              & 22nm                                                              & 22nm                                                              & 40nm                                                                    & 40nm                                                               \\ \hline
Input Precision (bit)                                                                               & 4/8                                                                       & 2/4/6/8                                                                                       & 4/8                                                               & 1/4/8                                                             & 1/2/4/8                                                           & 1-8                                                               & 1-8                                                                     & 1-8                                                                \\ \hline
Weight Precision (bit)                                                                              & 4/8                                                                       & 4/8                                                                                           & 4/8                                                               & 2/4/8                                                             & 2/4/8                                                             & 1-8                                                               & 4/8                                                                     & 4/8                                                                \\ \hline
Computing Mode                                                                                      & current domain                                                            & current domain                                                                                & charge domain                                                     & current domain                                                    & current domain                                                    & charge domain                                                       & current domain                                                          & charge domain                                                      \\ \hline
Multi-Bit Weight Processing                                                                         & digital shift-add                                                         & analog shift-add                                                                              & digital shift-add                                                 & digital shift-add                                                 & digital shift-add                                                 & digital shift-add                                                 & inherent shift-add                                                      & inherent shift-add                                                 \\ \hline
\begin{tabular}[c]{@{}c@{}}Average Circuit/Macro-Level \\ Energy Efficiency (TOPS/W)$\dagger$ \end{tabular}   & \begin{tabular}[c]{@{}c@{}}6.90@(8b,8b)$\star$ \end{tabular}   & \begin{tabular}[c]{@{}c@{}}41.67@(4b,8b)\\ (with sparse optimization) \end{tabular}         & \begin{tabular}[c]{@{}c@{}}9.26@(8b,8b) \end{tabular}    & \begin{tabular}[c]{@{}c@{}}3.60@(8b,8b) \end{tabular}    & \begin{tabular}[c]{@{}c@{}}4.72@(8b,8b) \end{tabular}    & \begin{tabular}[c]{@{}c@{}}6.53@(8b,8b) \end{tabular}    & \begin{tabular}[c]{@{}c@{}}12.18@(8b,8b) \end{tabular}         & \begin{tabular}[c]{@{}c@{}}14.47@(8b,8b) \end{tabular}        \\ \hline
\begin{tabular}[c]{@{}c@{}}Average System-Level \\ Energy Efficiency (TOPS/W)$\dagger$$\P$ \end{tabular}        & N/A                                                                       & \begin{tabular}[c]{@{}c@{}}9.40@(4b,8b) \end{tabular}                                & N/A                                                               & N/A                                                               & N/A                                                               & N/A                                                               & \begin{tabular}[c]{@{}c@{}}12.41@(4b,8b) \end{tabular}         & \begin{tabular}[c]{@{}c@{}}12.92@(4b,8b) \end{tabular}        \\ \hline\hline
\end{tabular}
}
\vspace{-1ex}
\begin{flushleft}
\scriptsize
$\dagger$: Scaled to 40nm, assume energy $\propto$ (Node)$^{2}$. 
$\P$: Based on CIFAR10-ResNet18. 
$\star$:@($x$b,$y$b): $x$-bit input, $y$-bit weight.\\
\vspace{-3ex}
\end{flushleft}
\vspace{-3ex}
\end{table*}

We further benchmark the system performance of our proposed analog IMC dual designs using NeuroSim V1.4 \cite{peng2019dnn+}, an integrated framework for DNN inference on IMC-based hardware accelerator with the support for various device technologies. 
We selected two representative networks, i.e., VGG8 and ResNet18, and two datasets, CIFAR10 and ImageNet, for our analysis. 
We assume an H-tree structure for routing among modules in each hierarchy. The sub-array size is set to 128x128,
and the partial parallel model for 32 input parallelism is enabled
as illustrated in Fig. \ref{fig:arch_CurFe} and \ref{fig:arch_ChgFe}. 
Moreover, the ON/OFF ratio of FeFET is set to $10^{5}$ \cite{soliman2020ultra}.
Modifications have been made to  NeuroSim  to accommodate our proposed  architectures.

In evaluating  DNN inference accuracy, we consider the impact of ADC resolution and input/weight precision,  as well as 
device variations under the CurFe and ChgFe architectures, as analyzed in  Section \ref{sec:results}. 
Using the VGG8 network on the CIFAR10 dataset as an example, with a baseline accuracy of 92\%, the experimental results in Fig. \ref{fig:systemacc} show that 5-bit ADC is necessary to avoid significant accuracy loss. This finding aligns with the analysis reported in \cite{peng2019dnn+}. Consequently, the ADC precision is set to 5-bit in the subsequent performance analysis.
The  accuracy under ChgFe is slightly lower than that under CurFe, consistent with  Fig. \ref{fig:MACout}, but  this discrepancy is  acceptable.
Even at 4-bit input/weight precision,  accuracy with 5-bit ADC under ChgFe is less than 0.5\% compared to CurFe.

Next, we evaluate the performance of the ResNet18 network on CIFAR10 and ImageNet. 
The system-level performance of our proposed dual designs, considering different input/weight precision, is shown in Fig. \ref{fig:systemperfm}. 
Consistent with  circuit-level results, ChgFe exhibits higher system energy efficiency than CurFe at the same precision level.
However, the throughput in ChgFe is lower than that in CurFe due to  longer time required for  MAC operation in ChgFe.
The system area costs are similar for both architectures. 
Furthermore, for the case of 4-bit input/weight precision, Fig. \ref{fig:systemeachlayer} provides a detailed breakdown of dynamic energy consumption and latency for each layer of ImageNet-ResNet18 in CurFe and ChgFe.

\vspace{-2.5ex}
\subsection{Comparison With  State-of-the-Art Designs}
\label{sec:Comparison}
\vspace{-1ex}

Table \ref{table:evaluation} compares our proposed designs with the state-of-the-art analog IMC designs that enable  8-bit precision DNN inference. 
For a relatively fair comparison of energy efficiency, all  designs are scaled to 40nm  with MAC operations of 8-bit input and 8-bit weights by multiplying $\lambda^{2}$, where $\lambda$ is the ratio of the realistic technology node to 40nm.
It should be noted that \cite{yue202014} includes additional sparse optimizations. 
Our analysis shows that without considering sparse optimization, 
our  FeFET-based analog IMC designs achieves the highest energy efficiency at the circuit level for 8-bit input and weight precision, 
which is 1.56$\times$ and 2.22$\times$ higher than the latest SRAM \cite{su202116} and ReRAM \cite{hung20228} based analog IMC designs, respectively. 
At the system level, using the same CIFAR10-ResNet18 configuration, the system energy efficiency of our FeFET-based analog IMC as derived from the Neurosim framework, is 1.37$\times$ higher than the latest analog IMC based system realization \cite{yue202014}. 
This enhanced efficiency is primarily due to 
the inherent shift-add capability in our CurFe/ChgFe designs, which eliminates the need for additional hardware for the shift-add operation in multi-bit weight processing. 
While the energy efficiency of the CurFe architecture is slightly lower than that of the ChgFe architecture, CurFe exhibits better robustness against device variations, as detailed in Fig. \ref{fig:systemacc}.

%% file: 05_Conclusion.tex
\vspace{-2.8ex}
\section{Conclusion}
\label{sec:conclusion}
\vspace{-1.5ex}

In this paper, we present novel dual designs for FeFET-based high precision analog IMC with inherent shift-add capabilities. 
These designs  capitalize on  the  analog storage characteristics of FeFETs to establish a unique
FeFET-based IMC array paradigm.
This approach  not only offers partial MAC capabilities for each column, but also inherently integrates the shift-add process for 4-bit weights directly within the array.
As a result, we eliminate the need for additional digital or analog shift-add circuits typically required for multi-bit weight processing.
Our proposed FeFET based IMC paradigm is designed to support both 2CM/N2CM MAC, thus providing flexible support for 4-/8-bit weight data in 2’s complement format. 
We then  develop the CurFe and ChgFe designs to accommodate the analog domain IMC architecture in both the current mode and charge mode, respectively. 
Evaluation results at circuit and system levels indicate that our novel dual designs offer superior energy efficiency  compared to existing state-of-the-art analog IMC approaches.